\documentclass[aps,prl,twocolumn,showpacs,superscriptaddress]{revtex4-1}
\usepackage{amssymb}
\usepackage{graphicx}

\begin{document}

\title{Anderson Localization at the Subwavelength Scale and Loss
Compensation for Surface-Plasmon Polaritons in Disordered Arrays of Metallic
Nanowires}
\author{Xianling Shi}
\author{Xianfeng Chen}
\author{Boris A. Malomed}
\author{Nicolae C. Panoiu}
\author{Fangwei Ye}
\email{[]fangweiye@sjtu.edu.cn}
\date{\today}

\begin{abstract}
Using a random array of coupled metallic nanowires as a generic example of
disordered plasmonic systems, we demonstrate that the structural disorder
induces localization of light in these nanostructures at a
deep-subwavelength scale. The \emph{ab initio }analysis is based on solving
the complete set of 3D Maxwell equations. We find that random variations of
the radius of coupled plasmonic nanowires are sufficient to induce the
Anderson localization (AL) of surface-plasmon polaritons (SPPs), the size of
these trapped modes being significantly smaller than the optical wavelength.
Remarkably, the optical-gain coefficient, needed to compensate losses in the
plasmonic components of the system, is much smaller than the loss
coefficient of the metal, which is obviously beneficial for the realization
of the AL in plasmonic nanostructures. The dynamics of excitation and
propagation of the Anderson-localized SPPs are addressed too.
\end{abstract}

\pacs{73.20.Mf, 42.82.Et, 78.67Pt, 78.68.+m}
\maketitle

\affiliation{State Key Laboratory on Advanced Optical Communication Systems and Networks, Department of Physics and Astronomy, Shanghai Jiao
Tong University, Shanghai 200240, China}

\affiliation{State Key Laboratory on Advanced Optical Communication Systems and Networks, Department of Physics and Astronomy, Shanghai Jiao
Tong University, Shanghai 200240, China}

\affiliation{Department of Physical Electronics, School of
Electrical Engineering, Faculty of Engineering, Tel Aviv University,
Tel Aviv 69978, Israel}

\affiliation{Department of Electronic and Electrical
Engineering, University College London, Torrington Place, London
WC1E 7JE, United Kingdom}

\affiliation{State Key Laboratory on Advanced Optical Communication Systems and Networks,
Department of Physics and Astronomy, Shanghai Jiao Tong
University, Shanghai 200240, China}










When the size of photonic devices is reduced to the subwavelength scale, the
confinement and guiding of the electromagnetic energy is severely hampered
by the diffraction of optical fields. This limitation represents the main
roadblock on the way to the integration of photonic circuits at the
nanoscale level \cite{Maier:07, Barnes:03}. An effective way to overcome
this limitation is to employ surface-plasmon polariton (SPP) waves \cite%
{Raether:88,Zayats2005131,Ozbay13012006}, whose strong confinement at the
metallic surface and deep-subwavelength characteristic scale make it
possible to achieve a strong coupling between the optical fields and
nano-sized photonic structures. In this context, one of major goals of the
work with SPP-based nanodevices is to develop new techniques for precise
beam steering, optical switching, and field manipulations at the
subwavelength scale. A very promising approach towards this goal is to
employ arrays of metallic nanowires, alias \textit{plasmonic crystals} \cite%
{Bozhevolnyi:01,Baumberg:05,Tao:08,Vasic:12,Giessen:12,Kivshar1,Kivshar2},
where the optical coupling of SPPs propagating in adjacent nanowires is
controlled by dielectric properties of the embedding medium \cite%
{Ye:10,Ye:11, Kou:13}. In particular, the use of periodic arrays of
nanowires makes it possible to engineer the effective optical dispersion
with an unprecedented degree of flexibility \cite%
{Vasic:12,Giessen:12,Kivshar1,Kivshar2}.

In this context, a natural question is to what extent structural disorder,
which is inevitably introduced by nanofabrication, or maybe purposely built
into the system, affects physical properties of the plasmonic crystals and
thus limits the functionality of subwavelength plasmonic nanodevices. In
particular, it is well known that the structural disorder may profoundly
affect the spectrum of wave modes, the Anderson localization (AL) being,
perhaps, the most spectacular effect of that kind. This is a fundamental
wave phenomenon, which was firstly predicted in solid-state physics as the
localization of electron wave functions in disordered lattices \cite%
{Anderson:58}. It has later been established that the AL is a ubiquitous
effect that occurs in a multitude of settings in which waves interact with
disordered potentials, including light \cite{John:84,Segev:07,Lahini:08},
matter waves \cite{Aspect:08,Roati:08}, and sound \cite{Hu:08}. Disorder
effects are expected to be particularly important at the subwavelength
scale, including plasmonic systems similar to those investigated in this
work, as the coupling between the waves and the underlying disordered system
is enhanced at that scale. In this context, the AL of SPPs was predicted in
metal-dielectric percolation composites \cite{Shalaev:99}, and effects of
randomly located scatters on SPP guiding along the surface of gold films
were observed experimentally \cite{Bozhevolnyi:02}.

In this paper, we study the influence of the structural disorder on the
spatial distribution of the plasmonic field and its propagation in one- and
two-dimensional (1D and 2D) arrays of coupled metallic nanowires. Solving
the full system of the corresponding Maxwell equations (ME), we find that a
random distribution of radii of the nanowires leads to transverse spatial
localization of collective SPP excitations (plasmonic \textit{supermodes} of
the array). The characteristic spatial confinement of the plasmonic field
may be significantly smaller than the optical wavelength, $\lambda $, which
demonstrates that plasmonic structures can be employed to implement the
subwavelength AL of the electromagnetic field. To facilitate experimental
observation of such extreme localization of light, we also study the
feasibility of the compensation of optical losses by means of embedded gain
elements. Our analysis shows that the deep-subwavelength Anderson-localized
SPPs may be maintained at extremely low gain levels, or, as a matter of
fact, even without gain.

\begin{figure}[t]
\centering \centering
\includegraphics[width=8.0cm]{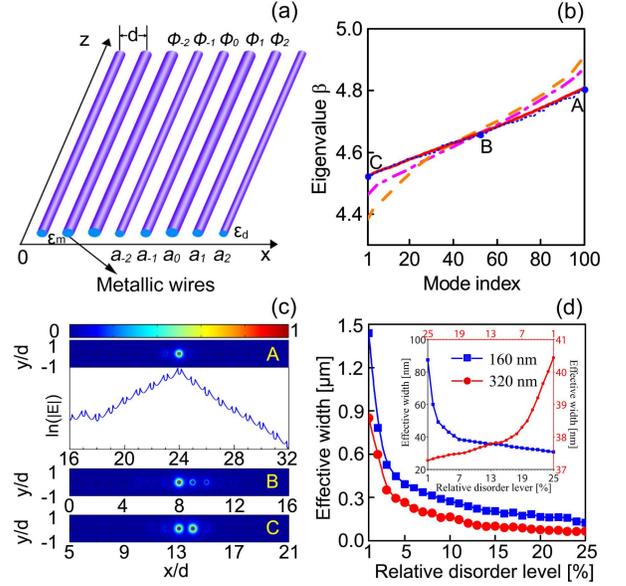}
\caption{(Color online) (a) A disordered plasmonic array.(b) The spectrum of
supermodes of the array, averaged over an ensemble of 100 randomness
realizations, produced by the CMT equations (\protect\ref{eGenCMEred}) in
the nearest-neighbor (dot-dashed line) and next-nearest-neighbor (dashed
line) approximations, as well as obtained from the full ME set (solid line).
The dotted line stands for the spectrum corresponding to one particular
realization of the randomness. Labels $A-C$ indicate the location of the
ALMs in (c). (c) Generic examples of electric-field intensity profiles of
ALMs in the array. The logarithmic plot shows the transverse profile of mode
($A$) at $y=0$. In (b) and (c), the array's spacing is $d=8a$,\ and the
randomness strength is $\Delta =10$\.{ }(d) The effective width of staggered
ALMs \textit{vs}. $\Delta $, as calculated with $d=4a$ and $d=8a$, for an
ensemble of 200 randomly composed plasmonic arrays. Other parameters in
(b)-(d) are $\protect\lambda =1.55~\mathrm{\protect\mu m}$, $a=40~\mathrm{nm}
$, $N=100$. Inset of (d) plots the effective width of ALMs at $\protect%
\lambda =0.628~\mathrm{\protect\mu m}$.}
\label{fig:geommode}
\end{figure}

We start by considering 1D arrays of $N$ coupled metallic nanowires, which
are oriented along the $z$-axis, being equally spaced (center-to-center) in
the transverse direction, $x$, by distance $d$, see Fig. \ref{fig:geommode}
(a). The structural disorder is introduced by fixing radii of the nanowires
in the array, with discrete coordinate $n$ , as $a_{n}=a+\delta _{n}$, where
$a$ is the average radius (we take $a=40~\mathrm{nm}$), and $\delta _{n}$ is
a random deviation. We assume that $\delta _{n}$ is uniformly distributed in
the interval of $[-\delta ,\delta ]$, with $\delta <a$, the level of the
disorder being characterized by $\Delta \equiv \delta /a$. The equal spacing
between nanowires makes the present setting different from fully random
plasmonic structures, \textit{e.g}., planar randomly distributed metallic
scatterers with random sizes \cite{Shalaev:99, Bozhevolnyi:02}; actually,
the nature of the disorder in the present system is similar to that
introduced by Anderson in his seminal work \cite{Anderson:58}.

Previous studies of the AL in systems of coupled waveguides were based on
the paraxial approximation for the propagation of electromagnetic waves,
chiefly because the relative variation of the refractive index in such
systems is small, hence the characteristic scale of the AL of light is much
larger than $\lambda $. However, the paraxial approximation is not valid for
plasmonic systems, where the relative variation of the refractive index is
large by definition, allowing, as we show below, the AL scale to be
significantly smaller than $\lambda $. This fact implies that the use of the
full system of three-dimensional (3D) ME is necessary. Thus, our analysis
starts \textit{ab initio}, solving the 3D ME system in the framework of the
COMSOL shell \cite{COMSOL}. In the simulations, a predefined triangular fine
mesh with a maximum-size element of 10 $\mathrm{nm}$ was used. The resulting
face mesh sweeps along the propagation direction of the nanowires with a
step of 500 $\mathrm{nm}$. Appropriate scattering boundary conditions were
used to mimic open boundaries. A convergence analysis was conducted to
ensure that the results vary within tolerable errors.

We set the permittivity of the dielectric background material to be $%
\epsilon _{d}=12.25$, which corresponds, \textit{e.g.}, to Si or GaAs, and
use the Drude model to describe the permittivity of the metal, $\epsilon
_{m}=1-\omega _{p}^{2}/\left[ \omega (\omega +i\nu )\right] $. We assume
that the nanowires are made of silver, with plasmon and damping frequencies $%
\omega _{p}=13.7\times 10^{15}~\mathrm{rad\cdot s^{-1}}$ and $\nu =2.7\times
10^{14}~\mathrm{rad\cdot s^{-1}}$\cite{Ordal:85}.

To gain better insight into the physics of the SPP localization, we compared
results produced by the ME system with those obtained from the paraxial
model for the propagation of SPPs in the disordered plasmonic arrays, based
on the coupled-mode theory (CMT). A detailed derivation of the CMT model
\cite{Ye:10,Ye:13,SI} leads to the discrete Schr\"{o}dinger equation with a
long-range coupling:
\begin{equation}
i\frac{d\phi _{n}}{dz}+b_{n}\phi _{n}+{\displaystyle\sum_{j\geq 1}}\kappa
_{j}(\phi _{n-j}+\phi _{n+j})=0,  \label{eGenCMEred}
\end{equation}
where $b_{n}$ is the propagation constant of the mode associated with the $n$
-th nanowire. Apart from the $z$-dependent phase, the nonvanishing field
components of this mode, $\mathbf{e}_{r}$, $\mathbf{e}_{z}$, and $\mathbf{h}
_{\phi }$, depend only on the radial coordinate, $r_{\perp }$. In Eq. (\ref%
{eGenCMEred}), $\kappa _{j}$ is the coupling coefficient between nanowires
separated by discrete distance $j$, which can be calculated using fields of
the plasmon mode of the nanowire, $\mathbf{e}(r_{\perp })$ and $\mathbf{h}
(r_{\perp })$, and the distribution of the dielectric constant in the
plasmonic system, $\epsilon (\mathbf{r}_{\perp })$ \cite{Ye:10}. The
corresponding modal fields are $\mathbf{E}(\mathbf{r})=\sum_{n}\mathbf{e}
_{n}(r_{\perp })e^{i(\beta k_{0}z-\omega t)}$ and $\mathbf{H}(\mathbf{r}
)=\sum_{n}\mathbf{h}_{n}(r_{\perp })e^{i(\beta k_{0}z-\omega t)}$, where $%
\beta $ is the effective refractive index of the plasmon mode, and $%
k_{0}=\omega /c$ the wavenumber in vacuum at carrier frequency $\omega$.

Results typical for the disordered plasmonic arrays are shown in Figs. \ref%
{fig:geommode}(b) and \ref{fig:geommode}(c), where the spectrum of the
supermodes (the transmission band) and several representative modal field
profiles of AL modes (ALMs) are displayed, respectively. Here we only show
results for the first transmission band, as higher-order ones do not support
localized eigenmodes. Our simulations, based on both the CMT and full ME,
reveal that, as expected, for small disorder ($\Delta \lesssim 5\%$) all
\emph{generic} supermodes of the \emph{finite} plasmonic array feature
extended profiles, \textit{i.e}., the AL does not occur, as, for such a weak
disorder, the propagation localization length may be larger than the
system's size (for some particular realizations of the disorder, the AL does
occur even if $\Delta $ is as small as $1\%$; however, for the very weak
disorder the AL is not a generic feature of the plasmonic field). When the
disorder strength exceeds $\Delta \simeq 5\%$, two strongly localized modes
emerge at edges of the transmission band, as shown in Fig. \ref{fig:geommode}%
(c). As may be naturally expected, ALMs near the bottom of the band are
\textit{unstaggered}, in the sense that the phase of the longitudinal
component of the electric field, $E_{z}$, is constant across the array,
while the ALMs at the top of the band are \textit{staggered} ($E_{z}$ in
adjacent nanowires points in opposite directions). ALMs located in the
central part of the band feature a mixed structure, with parts of the mode
staggered, and other parts unstaggered. When $\Delta $ increases further,
additional supermodes become more localized and evolve into ALMs. For all
values of $\Delta $ at which the AL occurs, the ALMs at the edges of the
band are, typically, localized much stronger than near its center.

Surprisingly, the CMT equations provide a somewhat more accurate description
of the plasmonic supermodes when only the nearest-neighbor coupling is kept
in Eq. (\ref{eGenCMEred}). Nevertheless, a particularly large discrepancy
between the predictions of the CMT and 3D ME is observed for the modes at
the edges of the transmission band. The ME and CMT not only predict
significantly different values for the propagation constant of the
supermodes of the plasmonic array, but also the field profiles of the
supermodes produced by these two methods are in poor agreement (not shown
here). These finding clearly demonstrate the\ necessity of the use of the 3D
ME for modeling strongly coupled, high-index-contrast systems, such as our
plasmonic arrays, as the CMT yields a coarse approximation in this setting.

Figure \ref{fig:geommode}(d) presents the effective width of the ALMs,
defined as
\begin{equation}
w_{\mathrm{eff}}=\left\langle \left[ \frac{\int_{-\infty }^{\infty }|\mathbf{%
\ E}(x,y=0)|^{2}(x-x_{0})^{2}dx}{\int_{-\infty }^{\infty }|\mathbf{E}
(x,y=0)|^{2}dx}\right] ^{\frac{1}{2}}\right\rangle,  \label{width}
\end{equation}%
where $x_{0}\equiv \int_{-\infty }^{+\infty }|\mathbf{E}(x,y=0)|^{2}xdx/
\int_{-\infty }^{+\infty }|\mathbf{E}(x,y=0)|^{2}dx$ is the central
coordinate of the mode, $\langle \rangle $ stands for averaging over
multiple realizations of the randomness with the same degree of disorder,
and the electric field is obtained by solving the 3D ME. Naturally, the
width decreases with the increase of the randomness strength, asymptotically
reaching a constant value for high disorder levels. When this minimum width
is reached, the plasmonic field is localized around a single nanowire. It is
worthy to note that the width can become much smaller than the wavelength
even at rather low disorder levels. For a given randomness strength, the
modal width increases with the decrease of the separation between the
nanowires, because smaller spacing leads to stronger coupling between them,
making stronger randomness necessary to induce the AL.

An important result inferred from Figs. \ref{fig:geommode}(b) and \ref%
{fig:geommode}(d) is that statistical averaging over the ensemble of
disordered arrays converges rather fast, hence a relatively small number of
arrays with different realizations of the disorder need to be actually
considered, deviations between the results produced by particular
realizations being small. This observation significantly reduces the
required computational time and thus greatly simplifies the analysis. This
result is explained by a weak dependence of the mode's propagation constant
and coupling strength on the radius of the nanowires. Equally important is
the potential implication of this result for the design of AL-based
plasmonic nanodevices, as one may expect that their properties weakly depend
on the particular realization of the system randomness \cite{Mafi:12}.
\begin{figure}[t]
\centering \centering
\includegraphics[width=8cm]{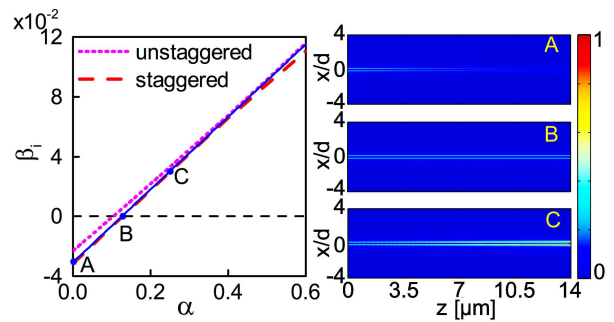}
\caption{(Color online) Left panel: the imaginary part of the propagation
constant, $\protect\beta _{i}$, \textit{vs}. the gain of the host dielectric
material, $\protect\alpha $, for staggered and unstaggered ALMs, averaged
over 100 randomness realizations. The permittivity of the host material and
metal are, severally, $\protect\epsilon _{d}=12.25+i\protect\alpha $ and $%
\protect\epsilon _{m}=-125.46-2.84i$ (at $\protect\lambda =1.55~\mathrm{%
\protect\mu m}$). Right panels: the propagation of the staggered ALM at
three different values of the gain for a specific randomness realization
(the solid line in the left panel): $\protect\alpha =0$ (A), $\protect\alpha %
=\protect\alpha _{\mathrm{cr}}=0.126$ (B), and $\protect\alpha =0.25$ (C).
The parameters are $d=320~\mathrm{nm}$, $N=20$, and $\Delta =15\%$.}
\label{fig:loss}
\end{figure}

The Ohmic loss in metallic nanowires causes decay of propagating ALMs, which
can make their observation a challenging task. A promising scheme to offset
the loss is to embed the array into a dielectric medium carrying optical
gain, provided,\textit{\ e.g}., by pumped quantum dots or wells \cite%
{Schaller:03}. Figure \ref{fig:loss} summarizes results produced by solving
the 3D ME for the loss characteristics of ALMs in the present setting, as
well as their formation and propagation in the presence of the gain. In the
simulations, we assumed that the metal's permittivity is $\epsilon
_{m}=-125.46-2.84i$, which corresponds to silver at $1550~\mathrm{nm}$, and
the permittivity of the embedding medium is $\epsilon _{d}=12.25+\alpha i$,
where $\alpha $ is the gain coefficient.

The most relevant quantity in this context is the imaginary part of the
ensemble-averaged modal propagation constant, $\beta _{i}$, as it directly
determines the loss. The dependence of $\beta _{i}$ of the staggered and
unstaggered ALMs on $\alpha $ is displayed in Fig. \ref{fig:loss}. Two
significant features are revealed by this figure. First, dependence $\beta
_{i}(\alpha )$ is almost linear. The reason for this is that the gain/loss
part of the permittivity of the metal and gain medium is much smaller than
the corresponding real part (especially in the metal, as stated above),
which means that the field profile of the ALMs remains almost unchanged as
one varies $\alpha $ (our numerical results directly confirm a weak
dependence of the modal profile on $\alpha $). As the effective loss of the
mode is given by a certain spatially weighted average of the imaginary part
of the permittivity over the nearly constant modal field profile, this
indeed implies that the corresponding loss coefficient depends on the
gain/loss almost linearly. Second, the gain coefficient, $\alpha _{\mathrm{cr%
}}$, at which the loss is compensated differs significantly from the loss
coefficient of the metal, being, quite surprisingly, much smaller than it.
This is explained by the fact that the mode does not distribute its field
evenly between the metallic (lossy) and gain regions. More specifically, the
simulations show that $\beta _{i}=0$ is achieved at $\alpha _{\mathrm{cr}%
}=0.126\ll 2.84$, which is more than 20 times smaller than the loss
coefficient. These findings are also illustrated by the propagation patterns
of ALMs, shown in the right panel of Fig. \ref{fig:loss} for three different
values of the gain coefficient: $\alpha =0<\alpha _{\mathrm{cr}}$, $\alpha
=\alpha _{\mathrm{cr}}$, and $\alpha =0.25>\alpha _{\mathrm{cr}}$.
\begin{figure}[t]
\centering \centering
\includegraphics[width=8cm]{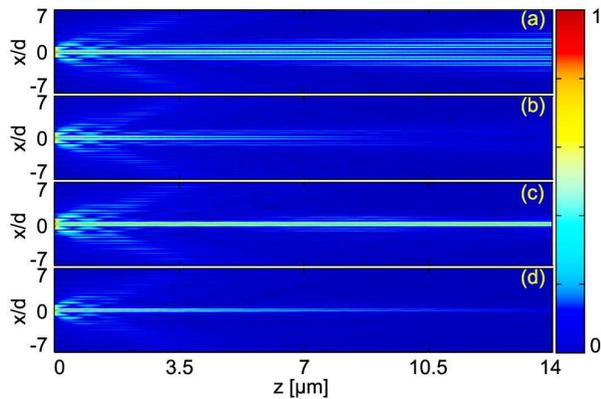}
\caption{(Color online) The propagation of a modulated Gaussian with
optimized parameters, defined by the following input: $E_{x}(x,y)=\sin [2%
\protect\pi (x-x_{0})/d]\exp [-(x-x_{0})^{2}/w^{2}]\exp (-y^{2}/w^{2})$,
which implies the excitation of a staggered ALM. Panels (a) and (b)
correspond to the periodic arrays, whereas (c) and (d) pertain to the
randomized ones. Metallic loss is not incorporated in (a) and (c), but is
included in (b) and (d). The parameters are $\protect\lambda =1.55~\mathrm{%
\protect\mu m}$, $d=320~\mathrm{nm}$, $x_{0}=d/2$, $w=d/2$, $N=20$, and $%
\Delta =15\%$.}
\label{fig:propag}
\end{figure}

In addition to the gain compensation, the excitation of ALMs, provided by an
appropriate input coupled into the system, is also an issue of critical
significance. Essential findings pertaining to this issue are summarized in
Fig. \ref{fig:propag}, which displays the excitation of ALMs by a modulated
Gaussian beam, whose initial width and input location in the array are
optimized using the field profile of the eigenmode provided by the above
analysis. The objective is to achieve the shortest ALM formation length,
defined as the distance required for the input beam to reshape itself into
an ALM. For the ALM to be observable in the setup in which the loss is not
compensated by the gain, the formation length should be shorter than the
characteristic modal decay length. Simulations presented in Figs. \ref%
{fig:propag}(c) and \ref{fig:propag}(d) clearly demonstrate that the
Gaussian, coupled into the disordered plasmonic array, evolves into an ALM
after passing just a few microns, a part of the input energy being shed off
in the form of radiation waves. In the course of the evolution, the beam
preserves its width, although its intensity exhibits an overall decrease if
realistic loss is included, see Fig. \ref{fig:propag}(d). By contrast,
significant beam diffraction is observed, over the entire propagation
distance, in arrays with vanishing randomness, as seen in Figs. \ref%
{fig:propag}(a) and \ref{fig:propag}(b). Thus, ALMs may be observed even in
the absence of the compensating gain, provided that the input profile is
properly adjusted. On the other hand, if the input significantly deviates
from the optimized shape, one should add the gain to make the ALM formation
length smaller than the modal decay length.

\begin{figure}[t]
\centering \centering
\includegraphics[width=8cm]{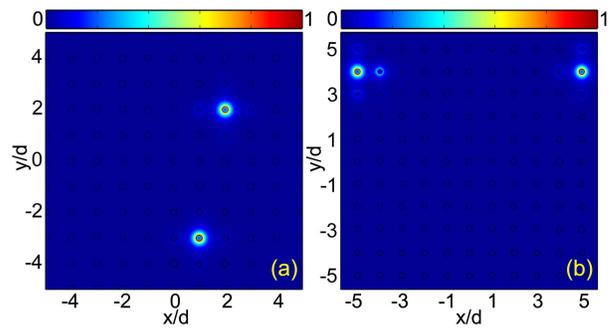}
\caption{(Color online) Examples of ALMs formed in a 2D disordered plasmonic
array. (a) and (b): Intensity of the electric field in bulk and surface
ALMs, respectively. In each panel, two modes are displayed, corresponding to
different realizations of the randomness. Parameters are $d=320~\mathrm{nm}$%
, $\Delta =15\%$, and $\protect\lambda =1.55~\mathrm{\protect\mu m}$.}
\label{fig:2D}
\end{figure}

The localization of SPPs is also possible in 2D disordered nanowire arrays.
Main features of the respective phenomenology are similar to those reported
above for the 1D setting, therefore we only briefly present them here. Two
representative examples of 2D ALMs are shown in Fig. \ref{fig:2D}, where
deep sub-wavelength confinement of the plasmonic field, in both transverse
directions, is clearly observed. In one case, the ALM is formed inside the
array, therefore we name it a bulk mode, whereas the other one is located at
the boundary of the array, and may be considered as a surface ALM. In both
cases, two ALMs formed under different randomness realizations are
displayed, which again shows that characteristics of AL in our disordered
system weakly depend on the particular realization. In particular, it is
observed that, even for a relatively weak disorder, the field is almost
entirely confined around a single nanowire. Our computations of the
propagation constant of the supermodes show that the predictions of the CMT
are still less accurate in 2D than in 1D.

In conclusion, by solving the complete set of the 3D Maxwell equations, we
have demonstrated that the AL\ (Anderson localization) of SPPs can be
achieved in 1D and 2D arrays of metallic nanowires with a varying degree of
the structural disorder. The characteristic localization length of these
plasmonic ALMs may be much smaller than the optical wavelength. We have
investigated the influence of the metallic loss and gain of the host medium
on the plasmonic ALMs, concluding that the loss is compensated by the gain
whose strength is much smaller than the loss rate of the metallic component
of the plasmonic array. These results suggest that experimental observation
of the ALMs (Anderson localization modes) is possible with the currently
available nanofabrication and experimental techniques.

It is worthy to note that the proposed settings can be readily extended to
the mid-IR and THz spectral regions by using other plasmonic systems, such
as arrays of graphene ribbons \cite{cbg12n,fra12n,cmt12acsn}. We also point
out that our approach, based on the full ME system, may be applied as well
to the subwavelength localization of atomic excitations, a phenomenon that
has been recently observed experimentally\cite{PRX}.

\begin{acknowledgments}
The authors thank D. Mihalache and S. K. Turitsyn for useful discussions. F. Ye acknowledges
financial support from the National Natural Scientific Funding of China (NSFC) (Grant No.
11104181) and Innovation Program of Shanghai Municipal Education Commission (Grant No. 13ZZ022).
X. Chen acknowledges financial support from NSFC (Grant No. 61125503). The work of N. C. Panoiu
was supported by the Engineering and Physical Sciences Research Council (Grant No. EP/J018473/1).
\end{acknowledgments}



\begin{thebibliography}{99}
\bibitem{} 
%
%

\bibitem{Maier:07} S.~A. Maier (Springer, Berlin, Germany, 2007).

\bibitem{Barnes:03} W.~Barnes, A.~Dereux, and T.~Ebbesen, {Nature} \textbf{\
424}, {824} ({2003}).

\bibitem{Raether:88} H.~Raether (Springer, Berlin, Germany, 1988).

\bibitem{Zayats2005131} A.~V. Zayats, I.~I. Smolyaninov, and A.~A.
Maradudin, Phys. Rep. \textbf{408}, 131(2005).

\bibitem{Ozbay13012006} E.~Ozbay, Science \textbf{311}, 189(2006).

\bibitem{Bozhevolnyi:01} S.~I. Bozhevolnyi, J.~Erland, K.~Leosson, P.~M.
Skovgaard, and J.~M. Hvam, Phys. Rev. Lett. \textbf{86}, 3008(2001).

\bibitem{Baumberg:05} T.~A. Kelf, Y.~Sugawara, J.~J. Baumberg,
M.~Abdelsalam, and P.~N. Bartlett, Phys. Rev. Lett. \textbf{95}, 116802
(2005).

\bibitem{Tao:08} A.~R. Tao, D.~P. Ceperley, P.~Sinsermsuksakul, A.~R.
Neureuther, and P.~Yang, Nano Lett. \textbf{8}, 4033 (2008).

\bibitem{Vasic:12} B.~Vasi\'{c} and R.~Gaji\'{c}, J. Opt. Soc. Am. B \textbf{%
\ 29}, 2964 (2012).

\bibitem{Giessen:12} C.~Bauer, G.~Kobiela, and H.~Giessen, Sci. Rep. \textbf{%
\ 2}, 681 (2012).

\bibitem{Kivshar1} A.~E. Miroshnichenko, B.~Lukyanchuk, S.~A. Maier, and
Y.~S. Kivshar, ACS Nano \textbf{6}, 837 (2012).

\bibitem{Kivshar2} I.~V. Iorsh, P.~A. Belov, A.~A. Zharov, I.~V. Shadrivov,
and Y.~S. Kivshar, Phys. Rev. A \textbf{86}, 023\,819 (2012).

\bibitem{Ye:10} F.~Ye, D.~Mihalache, B.~Hu, and N.~C. Panoiu, Phys. Rev.
Lett. \textbf{104}, 106802 (2010).

\bibitem{Ye:11} F.~Ye, D.~Mihalache, B.~Hu, and N.~C. Panoiu, Opt. Lett.
\textbf{36}, 1179 (2011).

\bibitem{Kou:13} Y.~Kou, F.~Ye, and X.~Chen, Opt. Lett. \textbf{38},
1271(2013).

\bibitem{Anderson:58} P.~W. Anderson, Phys. Rev. \textbf{109}, 1492 (1958).

\bibitem{John:84} S.~John, Phys. Rev. Lett. \textbf{53}, 2169 (1984).

\bibitem{Segev:07} T.~Schwartz, G.~Bartal, S.~Fishman, and M.~Segev, {Nature}
\textbf{446}, {52} ({2007}).

\bibitem{Lahini:08} Y.~Lahini, A.~Avidan, F.~Pozzi, M.~Sorel, R.~Morandotti,
D.~N. Christodoulides, and Y.~Silberberg, Phys. Rev. Lett. \textbf{100},
013\,906 (2008).

\bibitem{Aspect:08} J.~Billy, V.~Josse, Z.~Zuo, A.~Bernard, B.~Hambrecht,
P.~Lugan, D.~Clement, L.~Sanchez-Palencia, P.~Bouyer, and A.~Aspect, {Nature}
\textbf{453}, {891} ({2008}).

\bibitem{Roati:08} G.~Roati, C.~D'Errico, L.~Fallani, M.~Fattori, C.~Fort,
M.~Zaccanti, G.~Modugno, M.~Modugno, and M.~Inguscio, {Nature} \textbf{453},
{895} ({2008}).

\bibitem{Hu:08} H.~Hu, A.~Strybulevych, J.~H. Page, S.~E. Skipetrov, and
B.~A. Van~Tiggelen, {Nat. Phys.} \textbf{4}, {945} ({2008}).

\bibitem{Shalaev:99} A.~K. Sarychev, V.~A. Shubin, and V.~M. Shalaev, Phys.
Rev. B \textbf{60}, 16\,389 (1999).

\bibitem{Bozhevolnyi:02} S.~I. Bozhevolnyi, V.~S. Volkov, and K.~Leosson,
Phys. Rev. Lett. \textbf{89}, 186\,801 (2002).

\bibitem{COMSOL} $\mathrm{COMSOL~Multiphysics^{\circledR}}$
(http://www.comsol.com).

\bibitem{Ordal:85} M.~A. Ordal, R.~J. Bell, J.~R.~W.~Alexander, L.~L. Long,
and M.~R. Querry, Appl. Opt. \textbf{24}, 4493 (1985).

\bibitem{Ye:13} F.~Ye, D.~Mihalache, and N.~C. Panoiu (Springer, 2013), In
\emph{Spontaneous Symmetry Breaking, Self-Trapping, and Josephson
Oscillations}; Malomed B. A. Ed.; Progress in Optical Science and Photonics;
Vol.1; Chap. Sub-wavelength Plasmonic Solitons in 1D and 2D Arrays of
Coupled Metallic Nanowires.

\bibitem{SI} See Supplemental Material for a detailed derivation of the
coupled-mode equations.

\bibitem{Mafi:12} S.~Karbasi, C.~R.~Mirr, P.~G.~Yarandi, R.~J.~Frazier,
K.~W.~Koch, and A.~Mafi, Opt. Lett. \textbf{37}, 2304 (2012).

\bibitem{Schaller:03} R.~D. Schaller, M.~A. Petruska, and V.~I. Klimov, J.
Phys. Chem. B \textbf{107}, 13765 (2003).

\bibitem{cbg12n} J.~Chen, M.~Badioli, P.~Alonso-Gonzalez,
S.~Thongrattanasiri, F.~Huth, J.~Osmond, M.~Spasenovic, A.~Centeno,
A.~Pesquera, P.~Godignon, A.~Z. Elorza, N.~Camara, F.~J.~G. de~Abajo,
R.~Hillenbrand, and F.~H.~L. Koppens, Nature \textbf{487}, 77 (2012).

\bibitem{fra12n} Z.~Fei, A.~S. Rodin, G.~O. Andreev, W.~Bao, A.~S. McLeod,
M.~Wagner, L.~M. Zhang, Z.~Zhao, M.~Thiemens, G.~Dominguez, M.~M. Fogler,
A.~H.~C. Neto, C.~N. Lau, F.~Keilmann, and D.~N. Basov, Nature \textbf{487},
82 (2012).

\bibitem{cmt12acsn} J.~Christensen, A.~Manjavacas, S.~Thongrattanasiri,
F.~H.~L. Koppens, and F.~J.~G. de~Abajo, ACS Nano \textbf{6}, 431 (2012).

\bibitem{PRX} J.~A.~Miles, Z.~J.~Simmons, and D.~D.~Yavuz, Phys. Rev. X
\textbf{3}, 031014 (2013).
\end{thebibliography}
\end{document}